%% file: ms.tex
\documentclass[10pt,aps,prb,twocolumn, nofootinbib]{revtex4-1}

\usepackage{natbib}
\usepackage{url}
\usepackage{amssymb,amsmath}
\usepackage{subfig}
\usepackage{graphicx}
\usepackage{color}
\usepackage{hyperref}
\usepackage{listings}
\hypersetup{
    bookmarks=true,         
    unicode=false,          
    pdftoolbar=true,        
    pdfmenubar=true,        
    pdffitwindow=false,     
    pdfstartview={FitH},    
    pdfauthor={Kevin Multani},     
    colorlinks=true,       
    linkcolor=black,          
    citecolor=black,        
}
\graphicspath{{figures/}}

\usepackage{siunitx}
\usepackage[utf8]{inputenc}
\usepackage{mwe}

\newcommand{\Abbref}[1]{Figure \ref{#1}}

\newcommand{\Tabref}[1]{Table \ref{#1})}
\begin{document}

\definecolor{dkgreen}{rgb}{0,0.6,0}
\definecolor{gray}{rgb}{0.5,0.5,0.5}
\definecolor{mauve}{rgb}{0.58,0,0.82}

\lstset{frame=tb,
  	language=Matlab,
  	aboveskip=3mm,
  	belowskip=3mm,
  	showstringspaces=false,
  	columns=flexible,
  	basicstyle={\small\ttfamily},
  	numbers=none,
  	numberstyle=\tiny\color{gray},
 	keywordstyle=\color{blue},
	commentstyle=\color{dkgreen},
  	stringstyle=\color{mauve},
  	breaklines=true,
  	breakatwhitespace=true
  	tabsize=3
}

\title{MBE Growth of Al/InAs and Nb/InAs Superconducting Hybrid Nanowire Structures} 
\author{Nicholas A. Güsken\textsuperscript{1,2,3}, Torsten Rieger\textsuperscript{1,2}, Benjamin Bennemann\textsuperscript{1}, Elmar Neumann\textsuperscript{4}, Mihail Ion Lepsa\textsuperscript{1,2}, Thomas Schäpers\textsuperscript{1,2}, Detlev Grützmacher\textsuperscript{1,2} \vspace{0.5cm}} 

\affiliation{\textsuperscript{1}\textit{Peter Grünberg Institute (PGI-9), Forschungszentrum Jülich, 52425 Jülich, Germany}\\
\textsuperscript{2}\textit{JARA-Fundamentals of Future Information Technology (JARA-FIT), Jülich-Aachen Research Alliance, Germany}\\
\textsuperscript{3}\textit{Present address: Department of Physics, Imperial College London, London, SW7 2AZ, U.K.}\\
\textsuperscript{4}\textit{Peter Grünberg Institute (PGI-8), Forschungszentrum Jülich, 52425 Jülich, Germany} }

\begin{abstract}We report on \textit{in situ} growth of crystalline Al and Nb shells on InAs nanowires. The nanowires are grown on Si(111) substrates by molecular beam epitaxy (MBE) without foreign catalysts in the vapor-solid mode. The metal shells are deposited by electron-beam evaporation in a metal MBE. High quality supercondonductor/semiconductor hybrid structures such as Al/InAs and Nb/InAs are of interest for ongoing research in the fields of gateable Josephson junctions and quantum information related research.
Systematic investigations of the deposition parameters suitable for metal shell growth are conducted. In case of Al, the substrate temperature, the growth rate and the shell thickness are considered. The substrate temperature as well as the angle of the impinging deposition flux are explored for Nb shells. The core-shell hybrid structures are characterized by electron microscopy and x-ray spectroscopy. Our results show that the substrate temperature is a crucial parameter in order to enable the deposition of smooth Al layers. Contrary, Nb films are less dependent on substrate temperature but strongly affected by the deposition angle. At a temperature of $\SI{200}{\degree C}$ Nb reacts with InAs, dissolving the nanowire crystal. Our investigations result in smooth metal shells exhibiting an impurity and defect free, crystalline superconductor/InAs interface. Additionally, we find that the superconductor crystal structure is not affected by stacking faults present in the InAs nanowires.
\end{abstract}

\maketitle

\input{Introduction}

\input{Experimental}
\input{Results}
\input{Conclusion}

\bibliographystyle{phjcp}

\bibliography{library}
\end{document}

%% file: Introduction.tex
\section{Introduction}

InAs nanowires are promising building blocks for two major present and future topics in research i.e. applications in quantum computing \cite{Nadj-Perge2010,Leijnse2013} and the realization of Majorana fermions \cite{Das2012,Lutchyn2010}. Combining application and fundamental research, these intersecting subjects can be tackled using
superconductor/semiconductor (SC/SM) hybrid systems. For instance, the coupling of gate-defined qubits at the nanowire ends via a superconductor has been proposed \cite{Leijnse2013}, but also quantum network structures \cite{Alicea2011}. Further, gateable nanowire-based Josephson junctions \cite{Abay2014,Doh2005,Baba2015} are in the focus of ongoing research.
SC/InAs systems have been predicted to provide the optimal environment for proximity-induced p-wave superconductivity, leading in presence of an external magnetic field to spinless quasi-particles in form of triplet state Cooper pairs within the semiconductor \cite{Lutchyn2010,Oreg2010}. Moreover, InAs is known to be easily proximitized if brought into an impurity-free contact with Nb or Al \cite{Lutchyn2010,Doh2005}. This means that a superconducting gap can be induced within the semiconductor. These two ingredients combined with a large g-factor \cite{Bjork2005} and Rashba spin-orbit coupling \cite{Liang2012,EstevezHernandez2010} are prerequisites for the appearance of Majorana fermions \cite{Das2012}.
However, a clean SC/SM interface of high transparency is crucial to circumvent ambiguous experimental results and in order to provide an optimized performance for future devices \cite{Krogstrup2015}. Defects and impurities at the SC/InAs interface lead to parasitic subgap states which can be detrimental for topological quantum information \cite{Takei2013}.\\MBE grown SC/InAs hybrid nanowire structures exhibiting a distinct superconducting gap have been reported so far only by Krogstrup et al. \cite{Krogstrup2015}. They used Al as superconductor and phase pure InAs NWs grown by the Au-catalyzed vapor-liquid-solid method on InAs (111)B substrates. In the mentioned report, the epitaxy of Al shells took place at a substrate temperature of about -$\SI{30}{\degree C}$ and planar (on the substrate) Al growth rates of $0.3-\SI{0.5}{\micro\meter h^{-1}}$. Here, the Al-lattice orientations [111] and $[11\bar{2}]$ have been observed \cite{Krogstrup2015}. In this communication we seek to extend the range of superconductors feasible for contacting to niobium. Furthermore, self-catalyzed InAs NW growth instead of Au-catalyzed growth is applied, paving the way for a Si compatible technology.\\
Au-catalyzed growth can lead to the incorporation of Au impurities \cite{Du013, Allen08,Bar-sadan2012} and the creation of deep level defects which deteriorate the electrical and optical properties and are thus not compatible with Si technology \cite{Brotherton1980}. InAs NWs grown via the vapor-solid (VS) mode circumvent these obstacles, though they exhibit a strong polytypism. This manifests itself in a frequent switching between the zincblende (ZB) and the wurtzite (WZ) crystallographic phases, resulting in a high number of stacking faults.\\
In this paper, in-situ growth of the superconductor Al on VS grown InAs NWs using MBE is investigated. Our study extends the findings from metal deposition on phase pure wires \cite{Krogstrup2015} to NWs with alternating ZB and WZ phases. We find that the crystal orientation of the deposited metals is not affected by the occurrence of stacking faults within the InAs, i.e. the polytypism. 
Further, it is observed that the temperature has a major influence on the morphology of the NW shell. This is expected as for low temperatures the diffusion length of the Al adatoms is diminished \cite{Krogstrup2015}. A degassing procedure was developed in order to evaporate the As film formed after InAs NW growth, preventing from the formation of an AlAs interlayer. Our studies result in an impurity free and crystalline Al/InAs interface. 
Additionally, the deposition of niobium is analyzed highlighting the growth angle as the crucial parameter in order to obtain smooth and crystalline Nb shells. Nb possess the highest transition temperature and critical field of all elemental superconductors \cite{Peiniger1985}. Thus, the realization of Nb/InAs hybrid structures is of particular interest.

%% file: Experimental.tex
\section{Experimental}

The growth of SC/InAs hybrids was realized in a state of the art cluster, interconnecting several ultra-high-vacuum (UHV) chambers via transfer lines. This set up allows to conduct a multiple step growth process in situ. The InAs NWs have been grown in a III-V MBE chamber and the As capping which forms after growth has been degassed in a second As-free chamber. Subsequently, the Al and Nb SCs have been deposited in a metal MBE by electron-beam evaporation.\\
The NWs were grown in the $[\bar{1}\bar{1}\bar{1}]$ direction on Si(111) substrates in absence of foreign catalysts using the VS method. For the growth, clean substrates were specifically prepared. First, the native silicon dioxide of the Si substrate was removed by HF wet etching and DI-water. This was followed by treating the cleaned Si for $\SI{45}{s}$ with hydrogen peroxide \cite{Rieger2013}. This leads to the creation of pinholes inside a thin silicon dioxide layer, which serve as nucleation centers. After preparation, the substrates were transferred immediately into the load-lock chamber of the UHV-cluster.\\
The NWs were grown at a substrate temperature of $\SI{475}{\degree C}$ opening the In and As shutters for 1:20$\hspace{0.05cm}\mathrm{h}$. Here, an In effusion-cell and an As cracker-cell were used.
The In-growth rate was adjusted to $\SI{0.1}{\micro\meter h^{-1}}$ and the $\mathrm{As_4}$ beam equivalent pressure was set to about $\SI{3.00e-5}{Torr}$. The typical NW length was about $\SI{1.5}{\micro m}$ and its width about $\SI{120}{nm}$. After growth, a thin As film forms during the cool down, covering the wire. Before the Al deposition, a degassing procedure in a separate chamber was used to evaporate the arsenic and create a clean NW surface. First, the substrate was heated to $\SI{400}{\degree C}$ keeping this value for $\SI{10}{minutes}$, then it was heated to $\SI{450}{\degree C}$ using a $\SI{10}{minutes}$ ramp up. Here, the temperature was kept constant for $\SI{5}{minutes}$, followed by a consecutive cool down to $\SI{25}{\degree C}$. The metals Al and Nb were deposited via electron-beam evaporators at a partial pressure of $\mathrm{p_{Al}}=\SI{0.75e-5}{Torr}$ and $\mathrm{p_{Nb}}=\SI{1.5e-9}{Torr}$, respectively. The partial pressure was measured by a residual gas analyzer (RGA) and kept constant during growth via a feedback loop. The planar growth rates were $\mathrm{GR_{Al}}=\SI{0.97}{\micro \meter h^{-1}}$ and $\mathrm{GR_{Nb}}=\SI{0.26}{\micro \meter h^{-1}}$, respectively. The deposition time of the shells was $\SI{300}{s}$ for Al and $\SI{565}{s}$ for Nb, resulting in thicknesses of about $\SI{40}{nm}$ and $\SI{23}{nm}$, respectively.
Sub-zero temperatures were obtained by turning off all heat sources and loading the sample in the MBE chamber $\SI{15}{h}$ prior to growth. The temperature converged to a value of about $\SI{-6}{\degree C}$ after this time period. Even lower temperatures of about $\SI{-28}{\degree C}$ were reached by substituting the cooling water of the heater with cold vapor from liquid nitrogen.\\
The wire morphology as well as the crystal structure have been investigated using scanning electron microscopy (SEM) scanning transmission electron microscopy (STEM) as well as transmission electron microscopy (TEM) \cite{ErnstRuska-Centre2016}.

%% file: Results.tex
\section{Results \& discussion}

\subsection{Aluminum shell}
\begin{figure}[h!]
\center
{\includegraphics[width=8cm]{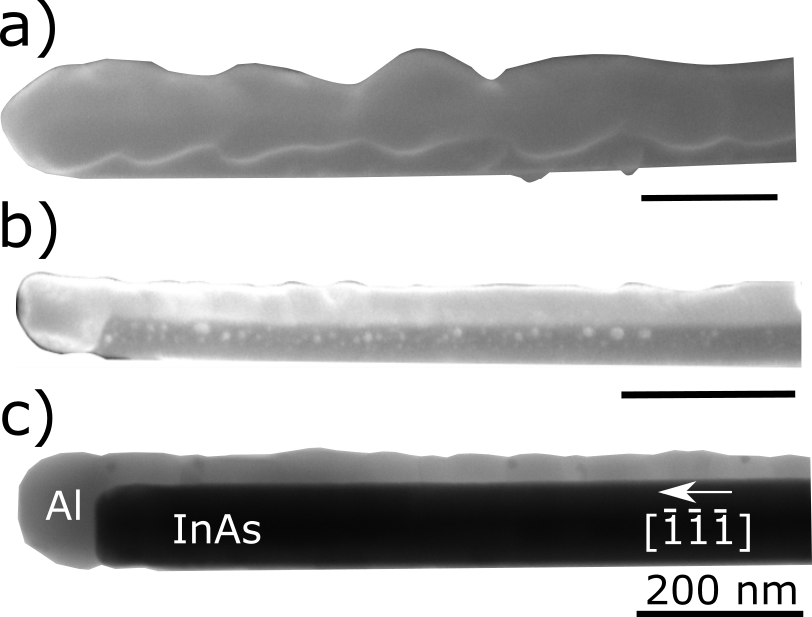}}
\caption{Side views recorded with SEM in a), b) and STEM in c) depicting InAs NWs covered with an Al half shell of about $\SI{40}{nm}$ thickness, grown at different substrate temperatures: a) $\mathrm{T_s}=\SI{42}{\degree C}$; b) $\mathrm{T_s}\approx\SI{-6}{\degree C}$; c) $\mathrm{T_s}\approx\SI{-28}{\degree C}$.}
\label{AlShellsmorpho}
\end{figure}

\begin{figure}[h!]
\center
{\includegraphics[width=8.6cm]{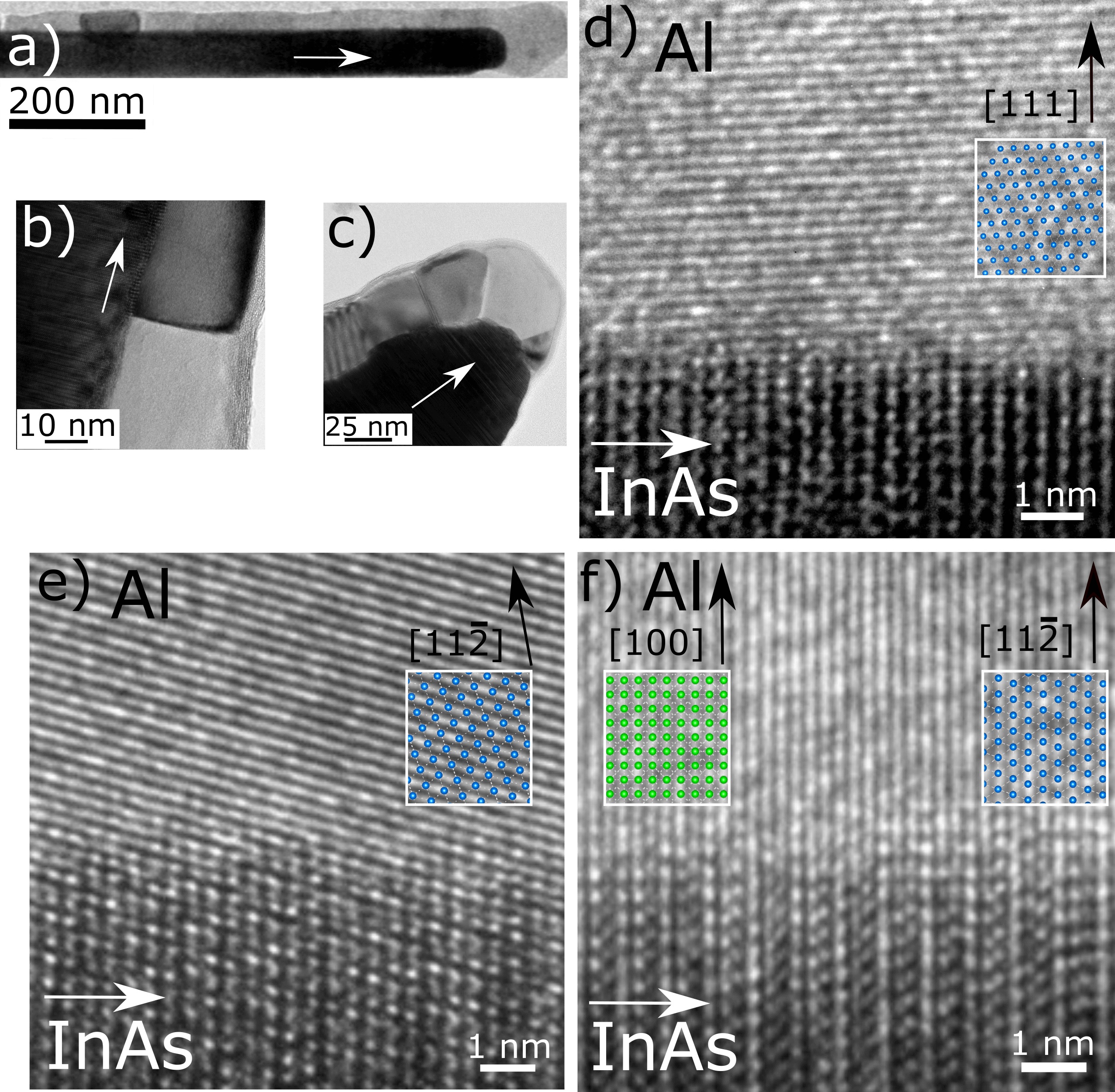}}
\caption{a)-c) TEM images of Al/InAs half shell structures. The metal deposition was conducted at $\mathrm{T_s}\approx\SI{-6}{\degree C}$. d)-f) InAs and Al crystal structure with superimposed lattice simulations of Al (insets) to identify the crystal arrangement. d) Crystalline Al on InAs oriented in the [111] direction. e) Crystalline Al in $[11\bar{2}]$ direction slightly deviating from wire normal as indicated by the black vector above the inset. f) Al in two possible orientations, indicated by the blue and green overlays. Al in $[11\bar{2}]$ direction but rotated by $\pi$ around the plane normal compared to (e), indicated in blue. Al in $\langle100\rangle$ direction, indicated in green. The InAs NW $[\bar{1}\bar{1}\bar{1}]$ growth direction is indicated in all images by the white arrows.}
\label{AlShellPolyCryst}
\end{figure}

Systematic studies of the growth parameters suitable to produce a smooth aluminum surface as well as an impurity and defect free Al/InAs interface were conducted. We found that deposition at a relatively high substrate temperature of $\SI{42}{\degree C}$ as depicted in \Abbref{AlShellsmorpho} {a)}, leads to a rougher morphology of the SC shell compared to depositions at lower temperatures, shown in \Abbref{AlShellsmorpho} {b), c)}. Growth at $\sim\SI{-6}{\degree C}$ results in a much smoother metal layer compared to the deposition at about $\SI{42}{\degree C}$. However, a lowering of the temperature from $\sim\SI{-6}{\degree C}$ to $\sim\SI{-28}{\degree C}$ did not change the growth morphology significantly as depicted in \Abbref{AlShellsmorpho} {c)}.
One possible explanation for the temperature dependence of the shell morphology is given by the growth model developed by Krogstrup et al., reporting a similar trend (supplementary information of reference \cite{Krogstrup2015}). The main idea in this approach is that the diffusion length of the impinging Al adatoms decreases with decreasing substrate temperature $\mathrm{T_s}$. This leads, for low $\mathrm{T_s}$, to the formation of small islands which cluster while the out-of-plane orientation is dominated by surface energy minimization. The direction of lowest energy, here the [111] orientation, grows dominantly at an early growth stage, i.e. for thinner layers.  At higher temperatures, the adatom diffusion length is larger, wherefore the islands exhibiting distinct lattice orientations grow bigger before they merge compared to the low temperature case. At a later growth stage, when the islands merge into clusters, the minimization of the grain boundary energy is predominant. Hence, the distinct grain orientation prevails, leading to a rougher metal shell morphology \cite{Krogstrup2015}.\\
\begin{figure}[h!]
\center
{\includegraphics[width=8.3cm]{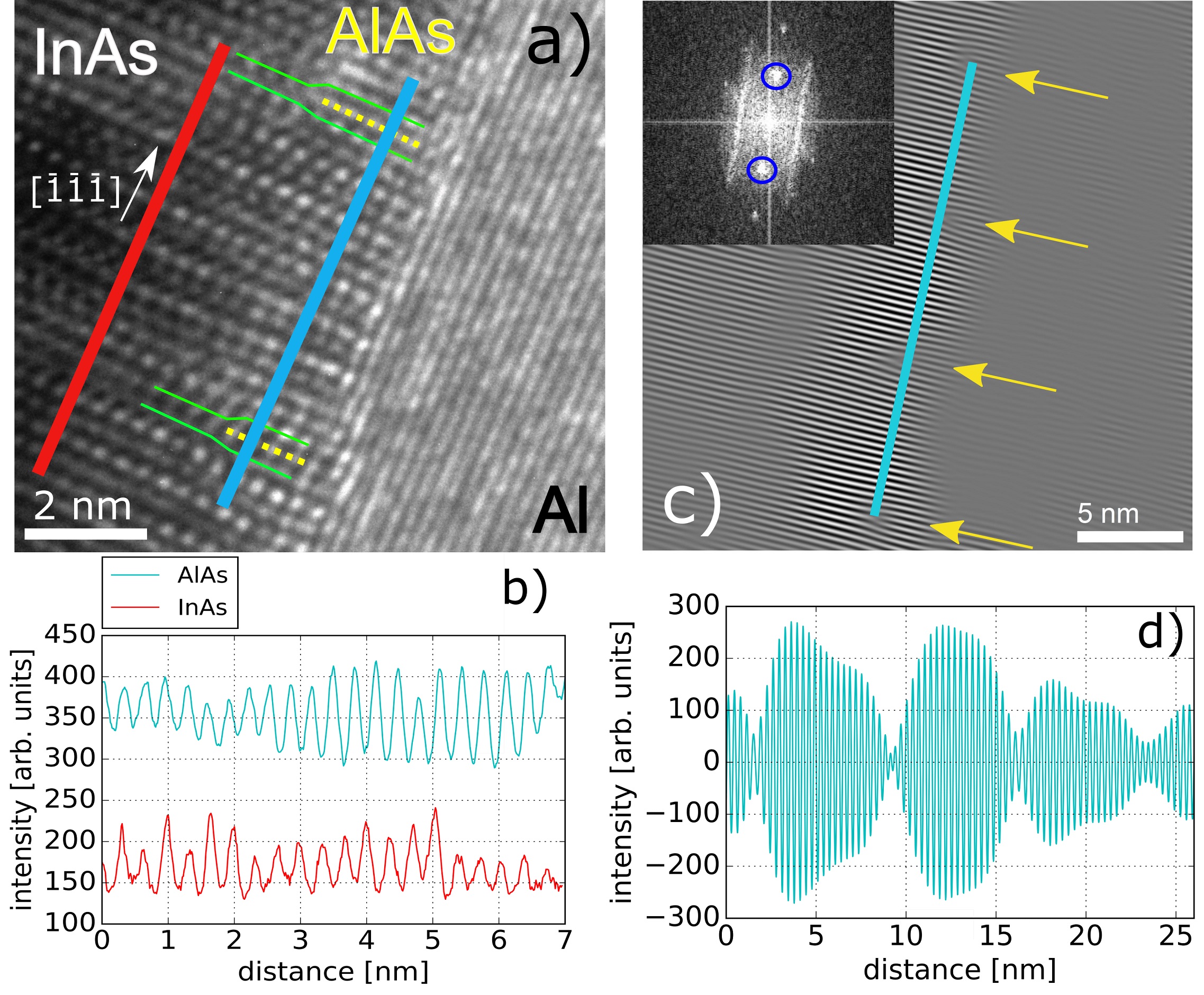}}
\caption{a) TEM image of an InAs NW with an AlAs interlayer and Al shell. Intensity scans along the red and turquoise colored lines are plotted in b). The yellow dotted lines between the green lines highlight the edge dislocations in AlAs. b) Profiles according to the line cuts of the AlAs and InAs to determine the lattice mismatch. c) Bragg filtered TEM image of an interface area to clearly indicate the distance between dislocations (marked with yellow arrows). The inset depicts the ZB $[\bar{1}\bar{1}\bar{1}]$ reflex used for the Bragg filtered TEM image. The blue line cut showing the distance between three dislocations is plotted in d).}
\label{PriorDegass}
\end{figure}
\Abbref{PriorDegass} {a)} shows an Al/InAs interface in which the InAs NW has not been treated with the degassing procedure introduced above prior to metal deposition. Thus, the InAs NW surface was covered with a thin layer of As. After Al deposition, an interlayer of a few nanometer thickness has formed, which we identified as AlAs. This assumption is based on estimations of the lattice mismatch between InAs and AlAs, explained below. The formation of an AlAs interlayer has been similarly observed by other groups \cite{Krogstrup2015, Ziino2013}. The AlAs interlayer leads to the formation of defects close to the interface.  As highlighted in \Abbref{PriorDegass} {a)} by yellow dotted lines, mainly edge dislocations are observed. Defects are expected to be detrimental for Majorana bound states \cite{Takei2013} and might degrade the induction of a superconducting gap.\\ The lattice mismatch between the InAs wire and the AlAs interlayer has been estimated in two ways. The first estimation of the lattice mismatch is based on intensity line profiles as exemplarily shown in \Abbref{PriorDegass} {b)}, from which the average lattice plane distances in the growth direction $\mathrm{d_{InAs,[\bar{1}\bar{1}\bar{1}]}}=0.339\pm\SI{0,007}{nm}$ and $\mathrm{d_{AlAs,[\bar{1}\bar{1}\bar{1}]}}=0.324 \pm \SI{0,003}{nm}$ were extracted. The lattice mismatch is then given by $\mathrm{m_{InAs/AlAs}}=(\mathrm{d_{InAs,[\bar{1}\bar{1}\bar{1}]}}-\mathrm{d_{AlAs,[\bar{1}\bar{1}\bar{1}]}})/\mathrm{d_{AlAs,[\bar{1}\bar{1}\bar{1}]}}$ and was determined to $\mathrm{m_{InAs/AlAs}} \approx 4.6\% $. In the second method, Bragg filtering was applied to visualize only the $(\bar{1}\bar{1}\bar{1})$ lattice planes (\Abbref{PriorDegass}{ c)}) from which the average dislocation distance $\mathrm{d_{Disloc.}}=7.197\pm \SI{0,002}{nm}$ was determined as indicated in \Abbref{PriorDegass} {c), d)}. The lattice mismatch is obtained via the relation $\mathrm{m'_{InAs/AlAs}=d_{InAs,[\bar{1}\bar{1}\bar{1}]}/d_{Disloc.}}$ and was estimated to $\mathrm{m'_{InAs/AlAs}} \approx 4.7\% $. The extracted values are smaller than expected for bulk ($\sim 7\%$). This can be attributed to the occurrence of stacking faults which cause a diminished lattice plane spacing in comparison to phase pure wires. Moreover, the thin AlAs interlayer ($\delta\approx\SI{2}{nm}$) is probably not completely relaxed, which leads to residual stress \cite{Biermanns2013}, resulting in a lower mismatch.\\
\begin{figure}[h!]
\center
{\includegraphics[width=8cm]{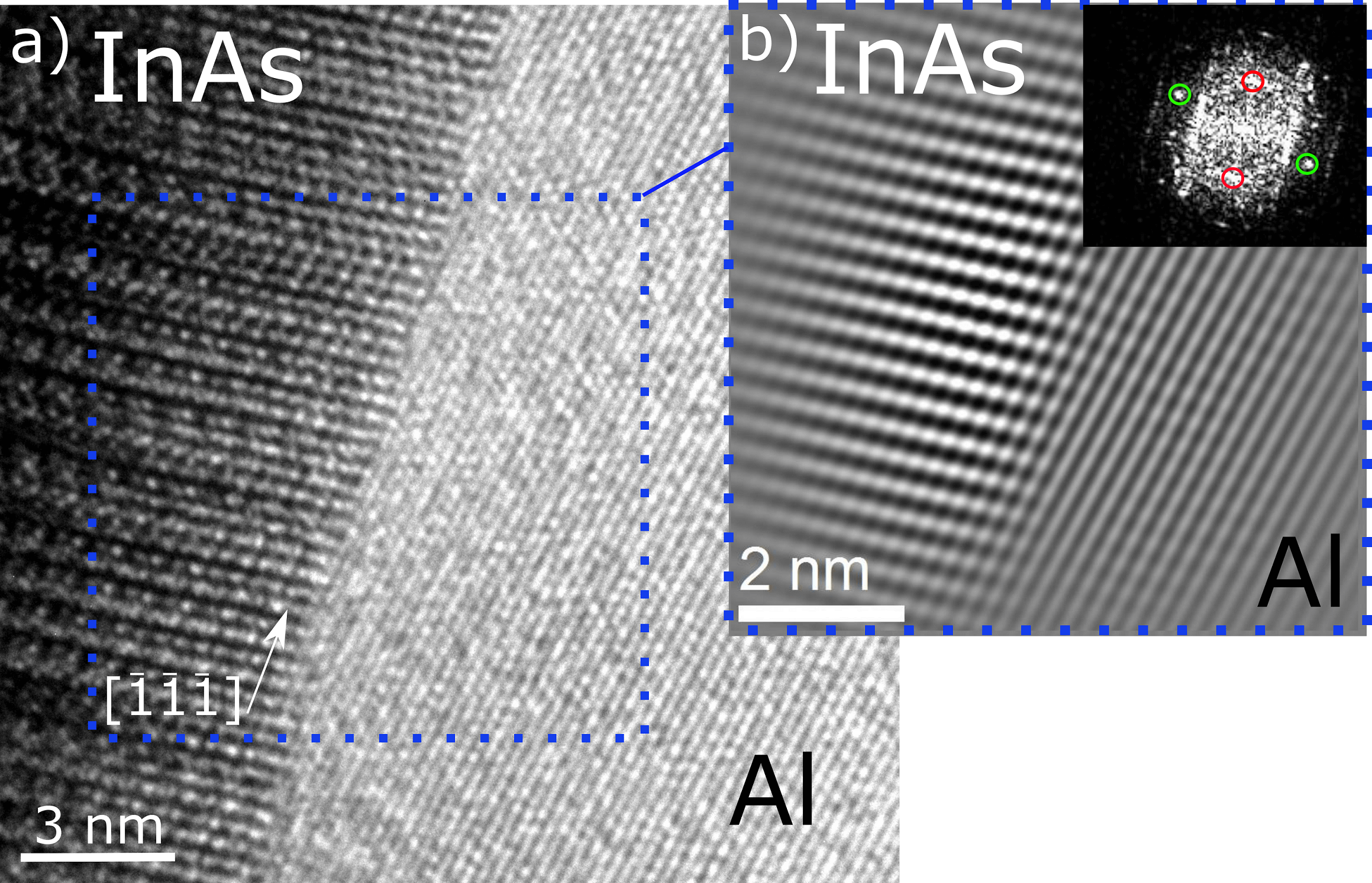}}
\caption{a) TEM image of an InAs NW showing the InAs/Al interface after As evaporation. b) Bragg filtered TEM image (fast-Fourier transformation (FFT) followed by an inverse FFT (IFFT)) depicting a section (blue dotted box) of the TEM image presented in a) to spotlight the absence of dislocations. The inset shows the FFT of the TEM image depicted in a). Here the ZB $[\bar{1}\bar{1}\bar{1}]$ reflexes of InAs are labeled red and the Al reflexes are labeled green, both are chosen for the IFFT.}
\label{InterfaceDegassed}
\end{figure}
\Abbref{InterfaceDegassed} depicts an Al/InAs interface after metal deposition with prior evaporation of the As layer formed after wire growth. An abrupt, impurity and defect free interface between the superconductor and the InAs nanowire is presented in \Abbref{InterfaceDegassed} {a)}. The Bragg filtered image depicted in \Abbref{InterfaceDegassed} {b)} exhibits, in contrast to \Abbref{PriorDegass} {b)}, no formation of dislocations, suggesting the absence of an AlAs interlayer. Thus, a defect and dislocation free Al/InAs interface can be assumed. We notice that no In clusters have been observed due to the annealing procedure for As evaporation.\\ The WZ/ZB polytypism of the InAs NWs does not affect the Al lattice structure, i.e. the phase switching characteristic of the NWs does not influence the crystal orientation of the SC shell. It means that no epitaxial growth of Al on InAs is present for NWs showing polytypism. This finding is an extension to the epitaxial growth for phase pure wires presented in literature \cite{Krogstrup2015}.
The Al shell grows polycrystalline, as illustrated in \Abbref{AlShellPolyCryst} {a)-c)}. However, long segments of several hundred nanometers with the same crystalline orientation are observed, as exhibited in \Abbref{AlShellPolyCryst} {a)}. The different lattice orientations are depicted in \Abbref{AlShellPolyCryst} {d)-f)}. The TEM images of the Al shell presented are superimposed by Al crystal simulations according to different Al directions in order to identify the lattice configuration. The simulations were created by using VESTA \cite{Momma2011}. We identify the low energy [111] direction depicted in \Abbref{AlShellPolyCryst} {d)}. Furthermore, the Al shell shown in \Abbref{AlShellPolyCryst} {e)} can be assigned to the $[11\bar{2}]$ direction, though its slightly deviating as indicated by the black arrow above the overlay. The lattice depicted in \Abbref{AlShellPolyCryst} {f)} corresponds either to the [100] or the $[11\bar{2}]$ direction. Though, the $[11\bar{2}]$ direction would be rotated by $\pi$ around the plane normal in comparison to the overlay presented in \Abbref{AlShellPolyCryst} {e)}. None of the images exhibits the transfer of the characteristic polytypism of the NW into the superconductor, i.e. a high crystalline quality of the Al shell is observed independent of the characteristics of the InAs core.

\subsection{Niobium shell}

\begin{figure}[h!]
\center
{\includegraphics[width=8.5cm]{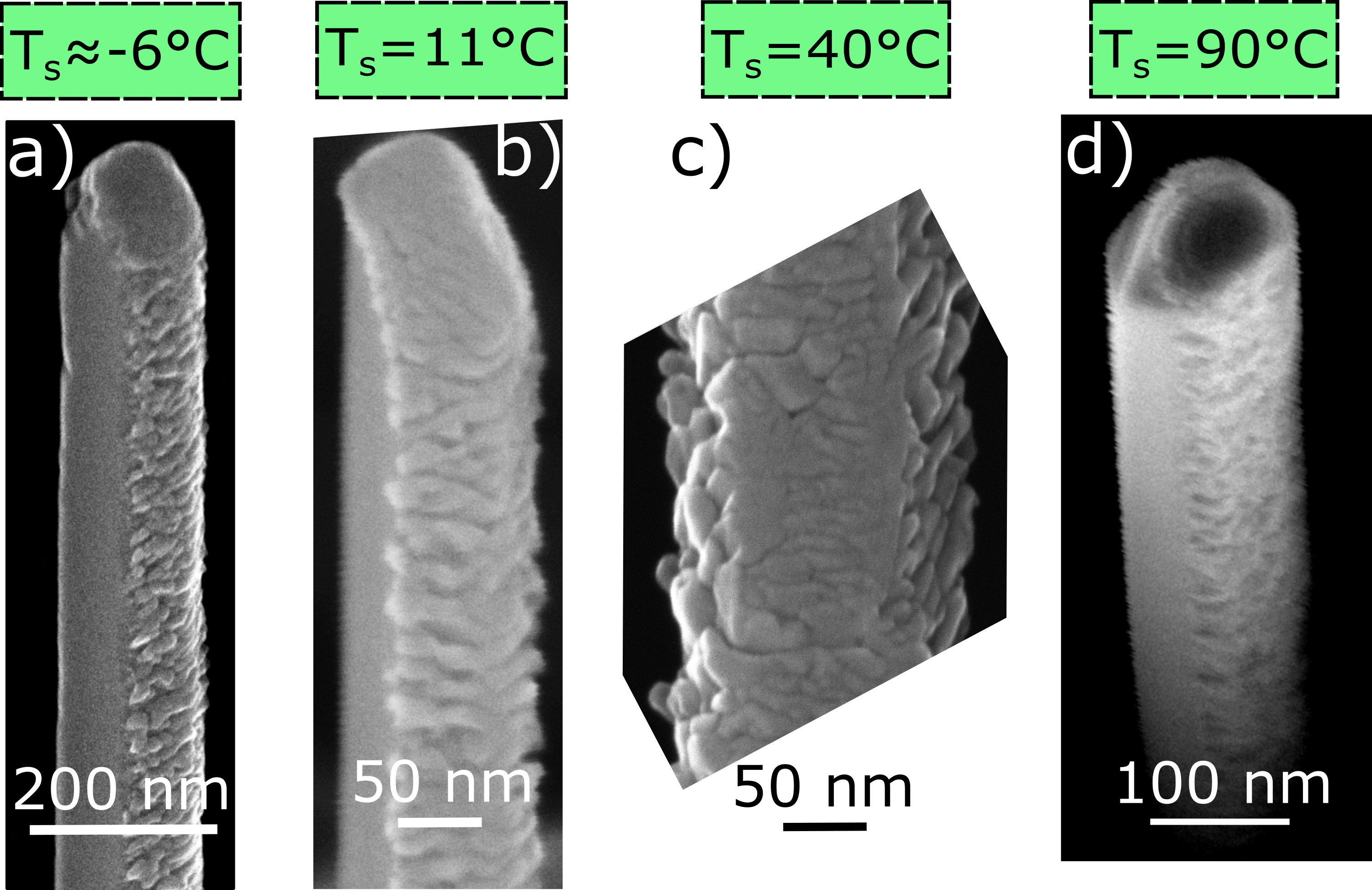}}
\caption{Nb shells on InAs NWs at different substrate temperatures $\mathrm{T_s}$. a), b), d) The shells depicted are about $\SI{23}{nm}$ thick; (c) The Nb shell is about $\SI{45}{nm}$ thick due to a longer deposition time.}
\label{NbShells}
\end{figure}
\begin{figure}[h!]
\center
{\includegraphics[width=8cm]{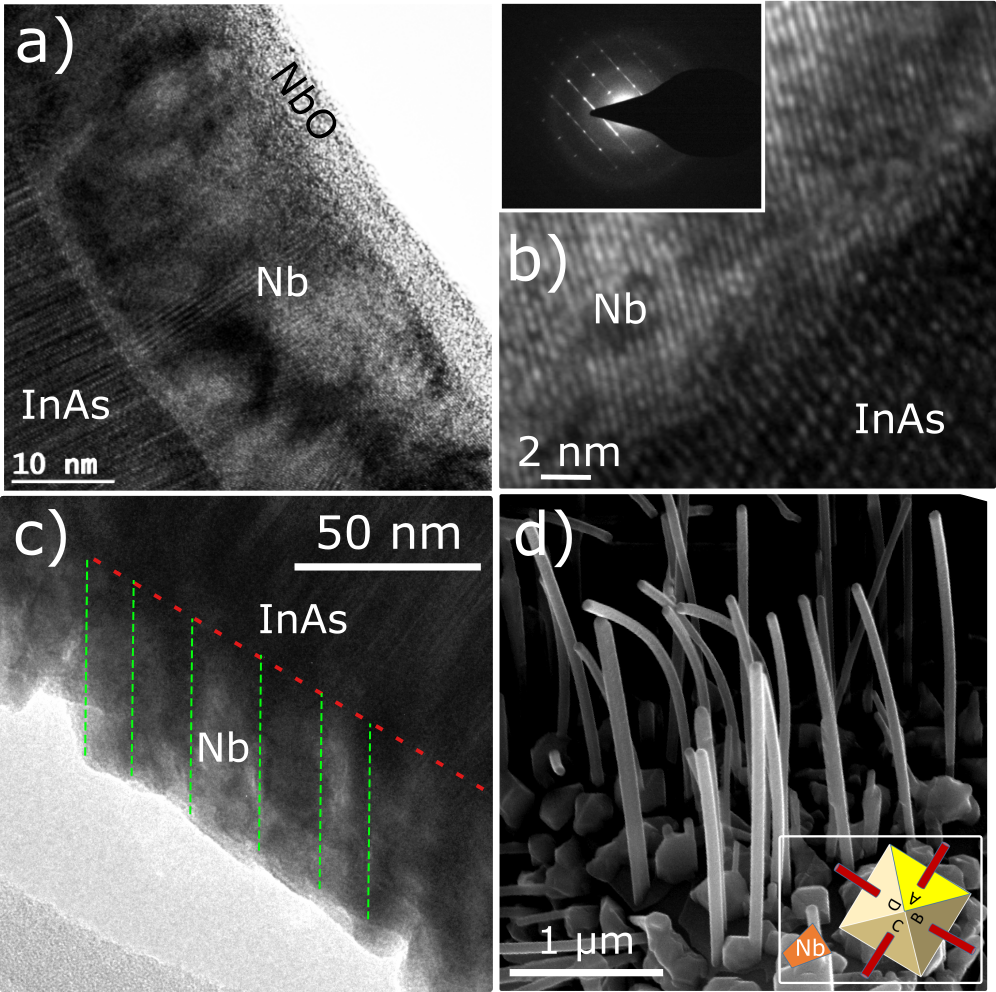}}
\caption{a) TEM image showing a Nb shell on an InAs NW covered by a thin NbO layer. The deposition angle is $\SI{87}{\degree}$. The thin white film at the InAs interface is due to defocusing effects of the TEM. b) TEM image of the interface, exposing the crystal structure. The inset depicts the corresponding electron diffraction pattern. c) Nb shell deposited under $\SI{25}{\degree}$. The colored lines are guidance for the eyes, pointing out the columnar-shaped growth mode. d) SEM side view of Nb/InAs NWs grown on a (111) side facet (facet A) of a pyramid (labeled yellow in the inset indicating the deposition geometry).}
\label{TEM}
\end{figure}

Niobium was deposited on vertically grown InAs NWs at a constant partial pressure of $\mathrm{p_{Nb}}=\SI{1.5e-9}{Torr}$. The substrate temperature $\mathrm{T_s}$ was varied between $\SI{-6}{\degree C}$ and $\SI{200}{\degree C}$. Two characteristic regimes of the Nb depositions are identified, below $\SI{90}{\degree C}$ and around $\SI{200}{\degree C}$. First, the observations made at $\SI{90}{\degree C}$ and below are presented. \\The SEM images of Nb-covered InAs NWs presented in \Abbref{NbShells} exhibit a streaked, columnar-shaped morphology almost independently of $\mathrm{T_s}$ within the temperature range examined.
This is contrary to the observations made for aluminum, for which $\mathrm{T_s}$ is the pivotal parameter. The behavior is possibly linked to the shorter diffusion length of Nb adatoms in comparison to Al. If the Nb adatoms are highly immobile they form mostly tiny clusters in close vicinity to their initial sites. As the deposition takes place under an angle of about $\SI{30}{\degree}$ between the electron-beam cell and the wire, the area behind these clusters is shadowed. This in turn might lead to the streaked columnar-like growth observed. In combination with the results obtained from the Al deposition, i.e. smoother shells for lower temperatures, the findings indicate the existence of an optimal diffusion length for each material suitable to form smooth and crystalline metal shells at a certain deposition angle. This length can be controlled via the substrate temperature. Within the temperature range investigated ($\sim\SI{-6}{\degree C}$ to $\SI{90}{\degree C}$) shells on vertical grown NWs remained rough, independently of the layer thickness (c.f. \Abbref{NbShells}). Hence, in case of Nb apparently, a substrate temperature above $\SI{90}{\degree C}$ is likely to allow for slightly smoother SC shells as the diffusion length is increased. Though, this temperature may already trigger chemical reactions between InAs and Nb, as explained below, wherefore a different technique was applied.\\
\begin{figure}[h!]
\center
{\includegraphics[width=8.5cm]{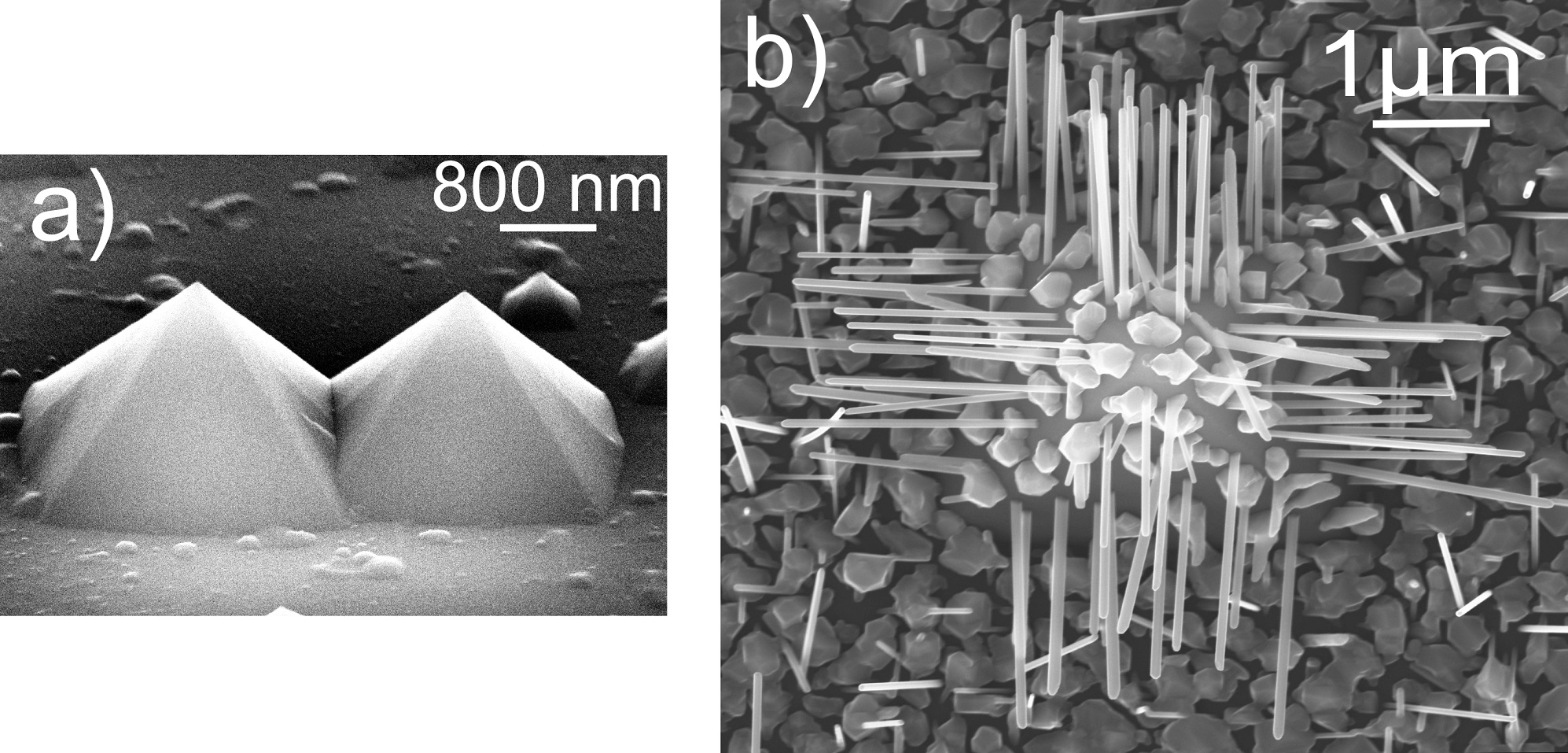}}
\caption{SEM pictures of silicon pyramids with and without InAs NWs. a) Side view of Si-pyramids with $\{111\}$ facets. b) Top view of InAs NWs on a Si pyramid.}
\label{Pyramids}
\end{figure}
Another approach enabling to overcome shadowing due to Nb clusters forming at the initial growth stage, is to alter the deposition angle between the metal flux and the side facets of the wire. For this reason, Si(100) substrates were etched in KOH solution obtaining Si pyramids with $\{111\}$ facets \cite{Rieger2016a} depicted in \Abbref{Pyramids} {a)}. This enables the growth of InAs NWs on the Si$\{111\}$ facets of the pyramids as shown in \Abbref{Pyramids} {b)}, which allows the deposition of Nb shells under different angles.\\
\begin{figure}[h!]
\center
{\includegraphics[width=8.5cm]{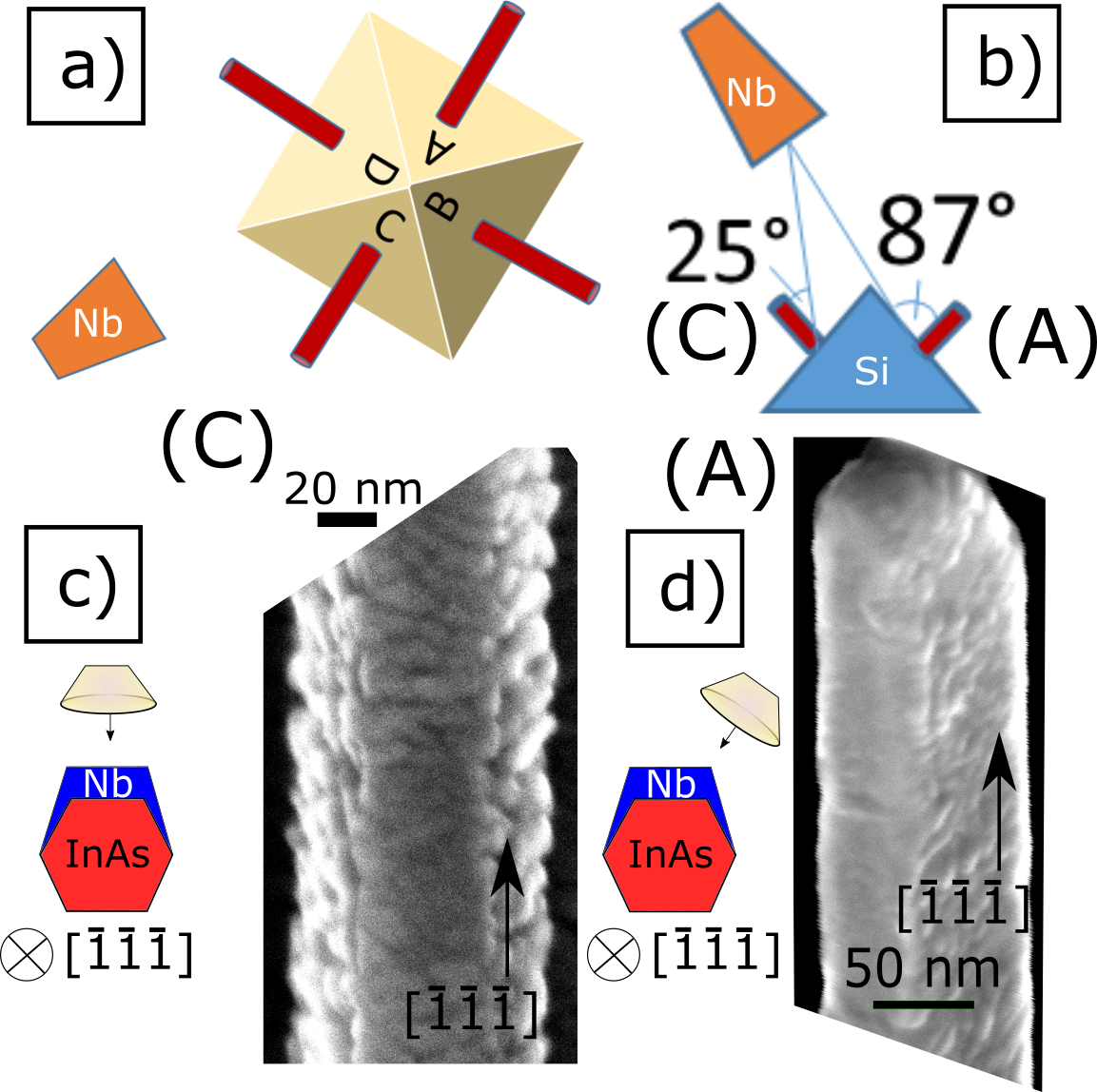}}
\caption{a), b) Schematic illustrating the orientation of the Nb crucible in respect to NWs (red) on pyramid facets. Wires from the pyramid facet C and A are shown exemplary in c), d) labeled (C), (A), respectively. The viewing direction onto the Nb coated InAs wire is sketched at the left-hand side of the NW images in c) and d).}
\label{NbShellPyram}
\end{figure}
The deposition geometry regarding the Nb flux and the side facets of the pyramids is sketched in \Abbref{NbShellPyram} {a),b)}. The Nb shell morphology illustrated in \Abbref{NbShellPyram} {c), d)} corresponds to the side facets C ($\sim \SI{87}{\degree}$) and A ($\sim \SI{25}{\degree}$) of the pyramids sketched in \Abbref{NbShellPyram} {a), b)}. Here, the substrate temperature is $\SI{40}{\degree C}$. The shell morphology exhibits a strong dependence on the growth angle. Deposition under an angle of $\SI{87}{\degree}$ between the Nb flux and the wire leads to a smooth and closed surface, as illustrated in \Abbref{NbShellPyram} {d)}. This contrasts with depositions conducted at an angle of $\SI{25}{\degree}$ which are still streaked and columnar-shaped as depicted in \Abbref{NbShellPyram} {c)}. The corresponding growth rates are estimated via SEM to $\mathrm{GR_{\SI{25}{\degree}}\approx\SI{0.3}{\angstrom s^{-1}}}$ and $\mathrm{GR_{\SI{87}{\degree}}\approx\SI{0.7}{\angstrom s^{-1}}}$, respectively. TEM images presented in \Abbref{TEM} confirm the different growth characteristics in dependence on the angle of deposition. \Abbref{TEM} {a)} depicts a smooth Nb shell on InAs of equal thickness, grown under an angle of about $\SI{87}{\degree}$. Here, apart from the InAs and the Nb, a NbO shell is visible. The Nb deposition conducted at an angle of $\SI{25}{\degree}$ results in a columnar-shaped morphology as shown in \Abbref{TEM} {c)}. 
The Nb/InAs interface illustrated in \Abbref{TEM} {b)} is impurity and defect free. The electron diffraction pattern shown in the inset of \Abbref{TEM} {b)} discloses the polycrystallinity of the Nb shell, as Debye-Scherrer rings occur. The streaks within the diffraction pattern are typical for InAs NWs exhibiting polytypism. Electrical investigations of Nb on InAs show that polycristalline growth does not strongly affect the superconducting properties \cite{Akazaki1991}.\\ \\
We observed that bending towards the coated side of the NWs occurs as shown in \Abbref{TEM} {d)}. 
Bending can be caused either by a difference in the thermal expansion coefficient or by a high lattice mismatch. The thermal expansion coefficients of InAs and Nb are $\SI{4.52}{\micro \degree C^{-1}}$ and $\SI{7.3}{\micro \degree C^{-1}}$ \cite{Ioffe}, respectively. Hence, the Nb shell will contract stronger than the InAs, when cooled to ambient temperature after deposition, which could cause the observed bending towards the coated side. However, no opposite bending occurs if the covered wires grown on a planar substrate are heated up from $\sim\SI{-6}{\degree C}$ to room temperature. No bending is seen, neither at $\SI{90}{\degree C}$ nor at $\SI{40}{\degree C}$ on vertically grown wires. In contrast, strong bending appears at $\SI{40}{\degree C}$ on wires grown on pyramids, depicted in \Abbref{TEM} {d)}. These finding suggests that the wires are bend due to strain induced by a lattice mismatch and not due to thermal contraction. However, the strength of the bending depends on the NW diameter, wherefore thicker wires are less affected and only slightly bend.\\

\begin{figure*}[]
\center
{\includegraphics[width=17.5cm]{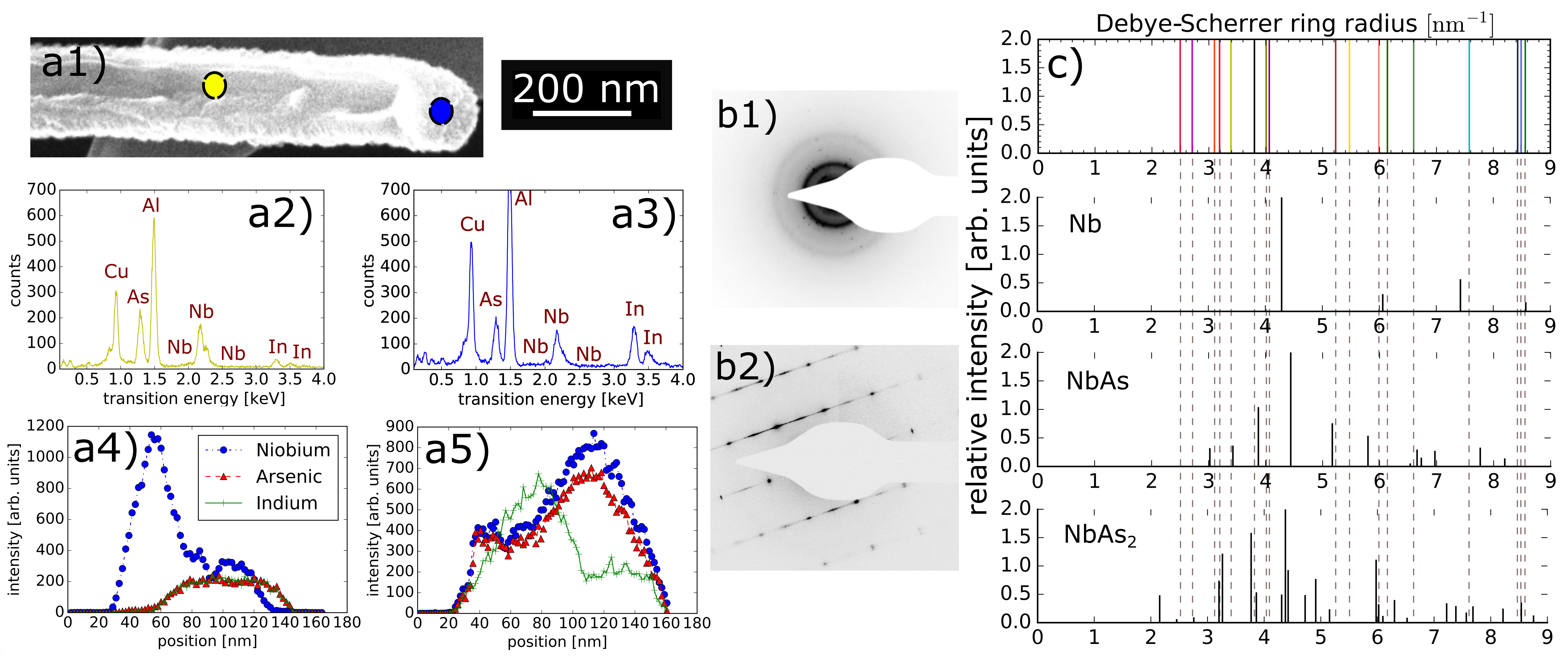}}
\caption{a1) Empty Nb shell stemming from Nb deposition on an InAs NW at $\SI{200}{\degree C}$ (scalebar on the right). a2), a3) EDX spectra recorded at the yellow dot (body) and blue dot (tip) in a1). The quantized atomic percentages of the tip spectrum shown in a2) as well as the body spectrum depicted in a3) are listed in \Tabref{QuantizedRatio}. a4, a5) EDX linescan (perpendicular to the wire axis) of a Nb/InAs half shell structure where the shell was deposited at $\SI{40}{\degree C}$ (a4)) and $\SI{200}{\degree C}$ (a5)). The legend for both plots is depicted in a4). b1)-b2) Electron diffractogram (inverted colors) from a Nb shell and an InAs NW, respectively. c) Debye-Scherrer ring radii extracted from b1). Nb compound patterns listed below for comparison \cite{ICSD}.}
\label{NbReaction}
\end{figure*}

Further, Nb depositions at a substrate temperature of $\SI{200}{\degree C}$ were conducted.
We observed that Nb reacts with InAs at this temperature.
The NWs have partly dissolved at $\SI{200}{\degree C}$, leaving empty Nb shells, as depicted in \Abbref{NbReaction} {a1)}. The morphology of the Nb has changed, exhibiting a dotted surface.
\begin{table}[]
\centering
\caption{Quantized atomic percentages extracted from the EDX scanned Nb shell deposited on an InAs NW at $\SI{200}{\degree C}$ (c.f. \Abbref{NbReaction}).}
\label{QuantizedRatio}
\begin{tabular}{l||l|l|l|l}
    & As & In & Nb  \\
    \hline
NW-Tip &   $43\%$ &  $30\%$  &  $27\%$   \\
NW-Body  &  $48\%$  &  $9\%$  & $43\%$  \\
\end{tabular}
\end{table}
The EDX spectra depicted in \Abbref{NbReaction} {a2), a3)} show that In is still present in the NW tip, but diminished in the body of the wire. The different atomic percentages of the respective species, i.e. As, In and Nb, between NW tip and body are listed in \Tabref{QuantizedRatio}. The high intensity Cu and Al peaks within the spectra stem from the TEM grid and the sample holder, respectively but are not of importance for the analysis. EDX linescans (normal to the wire axis) of Nb/InAs half shell structures for which Nb was deposited at $\SI{40}{\degree C}$ and $\SI{200}{\degree C}$ are depicted in \Abbref{NbReaction} {a4) and a5)}, respectively. Note that the shift of the peak positions is merely due to a flip of the wires. At a deposition temperature of $\SI{40}{\degree C}$ the As and In signal strongly overlap and the Nb shell signal is clearly distinct. At $\SI{200}{\degree C}$ in contrast, the In and As signal do not overlap anymore but a strong correlation between Nb and As is observed. This finding shows clearly that a chemical reaction between Nb and As was triggered at $\SI{200}{\degree C}$ during the deposition. 
\Abbref{NbReaction} {b1)} shows the electron diffraction pattern of a Nb shell deposited at $\SI{200}{\degree C}$. The occurrence of Debye-Scherrer rings is characteristic for polycrystalline materials. The absence of streaks demonstrates that the reaction with Nb at $\SI{200}{\degree C}$ was detrimental for the InAs NW. Diffraction patterns of vapor-solid grown InAs NWs show characteristic streaks due to their high number of stacking faults, as exemplified in \Abbref{NbReaction} {b2)} for comparison \cite{Rieger2015}.\\
The Debye-Scherrer diffraction pattern presented \Abbref{NbReaction} {c)} reveals that the Nb has reacted with As, forming new compounds. The obtained patterns show resemblance with the signatures of NbAs and $\mathrm{NbAs_2}$ but not with pure Nb.
We find that the substrate temperature sets a limit to the shell growth for two reasons. On the one hand, the deposition at $\SI{200}{\degree C}$ is detrimental as the Nb reacts with the InAs, dissolving the InAs crystal. This was equally observed for depositions at a substrate temperature of $\SI{400}{\degree C}$. On the other hand, the $\mathrm{Nb_x As_y}$ compounds formed are most likely not superconducting at ambient pressure anymore \cite{Zhang2015,Shen2016a} and therefore not suitable for SC/SM hybrid structures. 
\begin{figure}[h!]
\center
{\includegraphics[width=8.5cm]{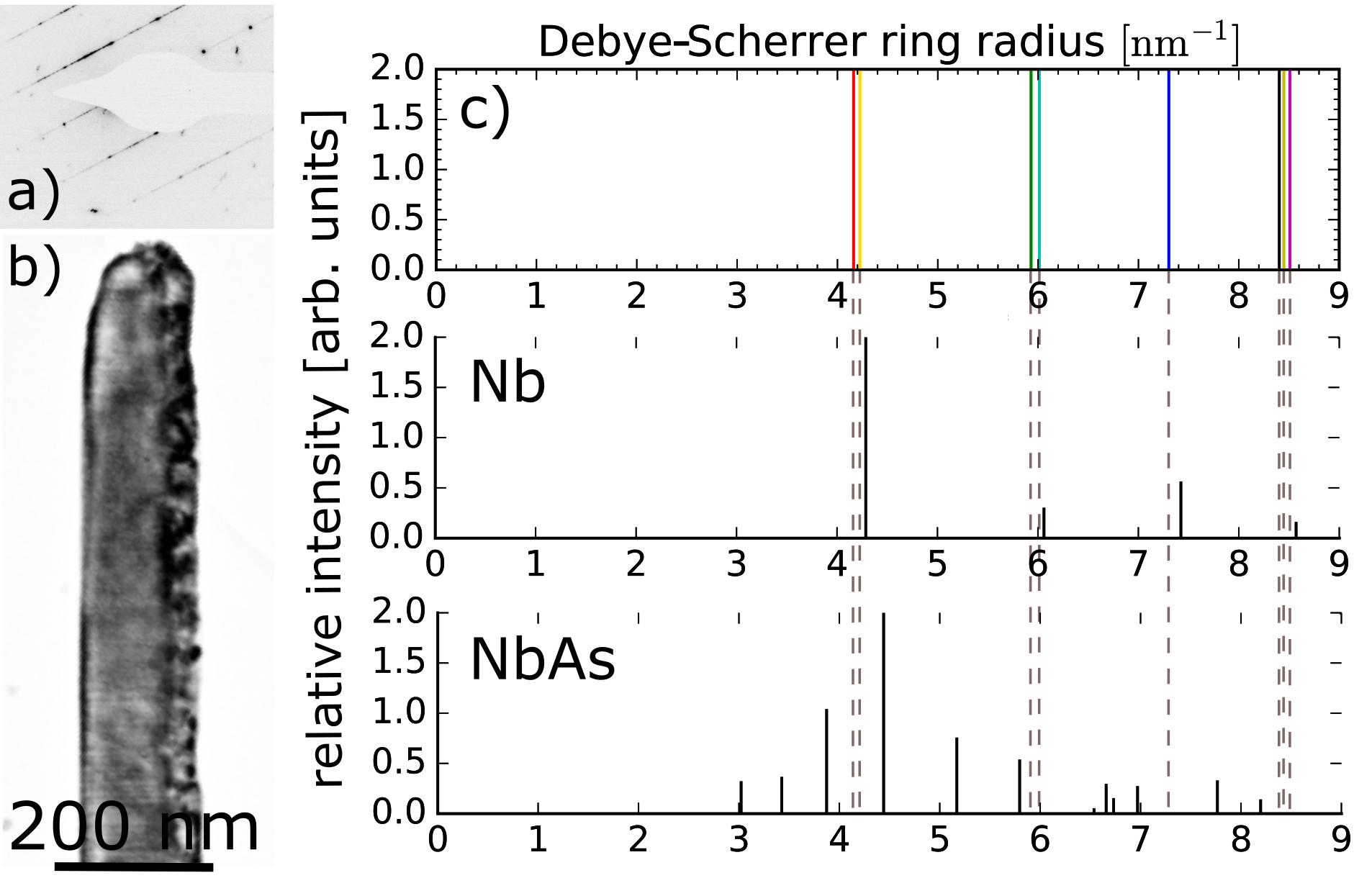}}
\caption{a) Electron diffraction pattern (inverted colors) of the Nb half shell covered InAs NW depicted in b). b) TEM image of  an Nb/InAs NW (wire on the left, shell on the right). The shell was deposited at $\SI{40}{\degree C}$ under an elevation angle of $\SI{87}{\degree}$. c) Extracted Debye-Scherrer ring radii for a Nb/InAs NW. Nb and NbAs Debye-Scherrer ring radii patterns are plotted for comparison \cite{ICSD}.}
\label{NbReaction40}
\end{figure}
However, at $\mathrm{T_s}=\SI{40}{\degree C}$ no reaction between Nb and InAs is observed. 
The electron diffraction pattern presented in \Abbref{NbReaction40} {a)} is clearly streaked, which proves the presence of an InAs NW. The side view of a niobium covered wire, presented in \Abbref{NbReaction40} {b)}, discloses the homogeneous morphology of the Nb shell. The Debye-Scherrer diffraction pattern depicted in \Abbref{NbReaction40} {c)}, indicates no additional compounds to Nb, suggesting the absence of a chemical reaction at $\SI{40}{\degree C}$.\\
Hence, the structures produced constitute promising candidates for further research at low temperature.

%% file: Conclusion.tex
\newpage
\section{Conclusion}

In conclusion, we produced self-catalyzed Al/InAs nanowire hybrid structures via MBE whereat the InAs NWs were grown within the vapor-solid mode. This report extends the investigations conducted on MBE grown SCs on phase pure InAs wires to SCs deposited on NWs exhibiting a high number of stacking faults. The hybrid structures disclose an impurity and defect free Al/InAs interface with a crystalline Al shell. A degassing procedure was developed to evaporate the As film which forms after the NW growth. The wires possess a closed and smooth Al shell which is suitable for further processing. The SC is polycrystalline, although segments of several hundreds of nanometers show monocrystalline Al shell growth. We observed that the Al crystal is oriented within different directions which formation can be attributed to boundary and surface strain energy minimization \cite{Krogstrup2015}. Our findings show that the crystal structure of the Al shell is not affected by the polytypism present in the InAs core. No transfer of stacking faults at present within the InAs wire into the SC deposit is observed, suggesting that the Al crystal orientation is not predominantly defined by the NW crystal structure. \\ 
In order to extend the material combinations at hand, the growth of Nb/InAs hybrids was investigated. We observed that the growth behavior is different from Al, which can be attributed to the different diffusion length of the adatoms. In case of Nb, the substrate temperature plays only a minor role. We found that the deposition angle, imposing a higher growth rate, changes the growth drastically. Smooth and coherent Nb shells can be produced by changing the angle between the Nb flux and the wire to $\SI{87}{\degree}$. The interface exhibits a crystalline structure and does not show any defects or impurities. The observation suggests a highly transparent interface suitable for low temperature investigations. Further, our findings revealed a chemical reaction between the InAs wire and Nb at a temperature of $\SI{200}{\degree C}$. The InAs crystal was dissolved, the arsenic reacting with the Nb.
As in the case of Al, the Nb crystal structure seems not to be influenced by stacking sequence within the NW.\\
This study is of importance to lay the groundwork for future quantum information technology devices based on superconducting hybrid structures. Crystalline SC shells and highly transparent SC/SM interfaces are crucial prerequisites within this field. As demonstrated, these requirements can be fulfilled using MBE growth. The realization of new material combinations such as Nb/InAs in addition to Al/InAs extends the temperature range and the critical fields possible for further research. \\

{\bfseries Acknowledgement}. The authors thank Christoph Krause for the support at the MBE and the team of the Helmholtz Nano Facility for the assistance concerning the sample processing.